\documentclass[11pt]{article}

\usepackage[margin=1in]{geometry}
\usepackage[T1]{fontenc}
\usepackage{microtype}
\usepackage{amsmath,amssymb}
\usepackage{booktabs}
\usepackage{graphicx}
\usepackage{algorithm}
\usepackage{algpseudocode}
\usepackage[numbers,sort&compress]{natbib}
\usepackage[hidelinks]{hyperref}

\title{\textbf{Anchor and Perturb: Lazy Agent Remediation by Exploration Injection}}
\author{Chengxi Zhong, Yongzhe Chang, Shiqing Yao}
\date{November 2026}

\begin{document}
\maketitle

\begin{abstract}
In MARL, teammates sharing a team reward propagated via $\partial Q_{\mathrm{tot}} / \partial Q_i$ can all receive credit for positive outcomes, even when some agents contribute near zero and never explore higher-return actions. Under standard uniform $\epsilon$-greedy exploration, concurrent exploratory noise across all $N$ agents can obscure trivial contributions from individual agents, driving QMIX into late-stage policy regression. Current remedies heavily rely on auxiliary models, intrinsic reward shaping, or pre-trained offline value anchors. Therefore, we propose \textbf{Anchor and Perturb (AnP)}, an online framework for remedying lazy agents without altering QMIX's network architecture, loss function, or task reward. Its automated variant, \textbf{AnP-Auto}, monitors a real-time \emph{Laziness Disparity Index} $\mathcal{L}_i(t)$ built from relative Q-value spreads and empirical action entropy. Once training reaches a point where lazy agents may surface while team win rate remains below a mastery ceiling, exceeding the $\mathcal{L}_i(t)$ threshold triggers a bounded rank-1 ($M=1$) exploration pulse with $\phi$-annealed decay ($\phi > 1$) on the lagging agent while holding all $N-1$ teammates strictly greedy ($\epsilon_{-i} = 0$). Following this, a $\phi$-ramped cooldown gradually releases the teammates into uniform exploration without major disruption to the replay buffer. As a proof of concept, within a tightly budgeted regime covering up to $4{,}500$ episodes, a heterogeneous 10-agent SMAC \texttt{MMM} benchmark across five paired random seeds revealed that AnP-Auto remedied vanilla QMIX's late-stage regression and improved terminal performance on all five seeds. The method raised the terminal 500-episode greedy win rate from $30.20\% \pm 4.28\%$ to $37.20\% \pm 4.55\%$ ($+7.00$ percentage points; paired $t = 3.65, p = 0.0217$) and evaluation return by $+0.71$ ($p = 0.0037$).
\end{abstract}

\section{Introduction}
\label{sec:introduction}

Value-factorisation methods under Centralised Training with Decentralised Execution (CTDE)---VDN~\cite{sunehag2018vdn}, QMIX~\cite{rashid2018qmix}, Weighted QMIX~\cite{rashid2020wqmix}, and QPLEX~\cite{wang2021qplex}---decompose a joint action-value function $Q_{\mathrm{tot}}$ into individual agent utilities, $Q_i$, trained from shared team reward. This approach makes scalable multi-agent control possible.

The problem is that relying on a single scalar team signal induces the \emph{lazy agent} problem~\cite{sunehag2018vdn}. Once active teammates contribute to most of the team return, agents with trivial contributions settle into passive, low-entropy behaviours that make reaching higher rewards unlikely. Under standard uniform $\epsilon$-greedy exploration, this worsens due to \emph{relative overgeneralisation}~\cite{panait2006biasing,wei2016lenient}. Simultaneous exploration noise across all $N$ agents renders transition dynamics non-stationary. Individual marginal contributions become trivialised, and QMIX gets trapped in a suboptimal equilibrium---one prone to late-stage stagnation and regression on heterogeneous tasks like SMAC \texttt{MMM}.

Existing optimisations either alter objectives or depend on restrictive algorithms. Intrinsic-motivation and auxiliary-model approaches---MAVEN~\cite{mahajan2019maven}, EITI/EDTI~\cite{wang2020influence}, EMC~\cite{zheng2021emc}, LAIES~\cite{liu2023laies}---introduce extra indicators: latent variables, curiosity bonuses, causal world models. They consume unnecessary resources and shift the Bellman target away from the true task reward. Sequential methods limit multi-agent non-stationarity differently, concentrating updates on a single agent in on-policy settings (HAPPO~\cite{kuba2022hatrpo}, A2PO~\cite{wang2023a2po}) or via Decentralised Sequential Exploration in offline-to-online fine-tuning, as in OVMSE (Hai Zhong et al., 2025)~\cite{zhong2025ovmse}. But OVMSE requires a pre-trained model and randomly chooses which agent to perturb at each step. Breaking lazy-agent equilibrium online without pre-training or altering the learner's loss and architecture remains an open problem.

We propose \textbf{Anchor and Perturb (AnP)}, an online rollout-scheduling framework that leaves QMIX's architecture, reward signal, and Bellman target unchanged. AnP works on a simple principle: to flag and recover a stalled agent, its teammates must stay in pure exploitation while the stalled agent explores. Our closed-loop formulation (\textbf{AnP-Auto}) avoids perturbing all $N$ agents uniformly. It monitors an alive-masked \emph{Laziness Disparity Index} $\mathcal{L}_i(t)$ that combines relative Q-value spread suppression with exponential action-entropy damping. Once training passes an early-exploration borderline and team win rate remains below a mastery ceiling, exceeding the $\mathcal{L}_i(t)$ threshold triggers a bounded rank-1 ($M=1$) exploration pulse on $\arg\max_i \mathcal{L}_i(t)$ with $\phi$-annealed decay ($\phi > 1$), while strictly anchoring remaining $N-1$ teammates to deterministic greedy execution ($\epsilon_{-i} = 0$). A smooth $\phi$-ramped cooldown then un-anchors teammates back to nominal exploration without distributional shock.

On the heterogeneous 10-agent SMAC \texttt{MMM} benchmark across five paired random seeds, AnP-Auto breaks vanilla QMIX's late-stage regression and improves terminal performance on every seed. Terminal 500-episode greedy win rate rises from $30.20\% \pm 4.28\%$ to $37.20\% \pm 4.55\%$ ($+7.00$ percentage points; paired $t = 3.65, p = 0.0217$). Evaluation return increases by $+0.71$ ($p = 0.0037$). Three contributions:
\begin{enumerate}
    \item We formulate the alive-masked Laziness Disparity Index $\mathcal{L}_i(t)$, a model-free online diagnostic that flags credit-starved agents directly from standard Q-values and action histories.
    \item AnP-Auto is an objective-preserving online exploration schedule uniting strict $N-1$ teammate anchoring, performance-gated $\phi$-annealed rank-1 pulses, and smooth post-pulse un-anchoring.
    \item Counterfactual paired-seed validation on SMAC \texttt{MMM} demonstrates statistically significant terminal win-rate and return improvements alongside closed-loop probe telemetry.
\end{enumerate}

\section{Related Work}
\label{sec:related_work}

\paragraph{CTDE Value Factorisation and Credit Assignment.}
Value-factorisation methods decompose a centralised joint action-value function $Q_{\mathrm{tot}}$ into decentralised agent utilities $Q_i$, satisfying the Individual-Global-Max (IGM) condition. VDN~\cite{sunehag2018vdn} enforces additive factorisation, whereas QMIX~\cite{rashid2018qmix} generalises this to state-conditioned monotonic mixing via non-negative hypernetworks. Subsequent architectures relax QMIX's monotonic projection bias through weighted Bellman targets (Weighted QMIX~\cite{rashid2020wqmix}) or duplex dueling advantage decompositions (QPLEX~\cite{wang2021qplex}). Actor-critic methods such as COMA~\cite{foerster2018coma} address credit assignment in parallel, using explicit counterfactual baselines. AnP operates solely at the rollout exploration interface without altering the mixing network, loss function, or Bellman target. As a result, it is orthogonal to these architectural extensions and integrates directly onto QMIX.

\paragraph{Lazy Agents and Relative Overgeneralisation.}
Shared team rewards induce the \emph{lazy agent} pathology~\cite{sunehag2018vdn}, where passive agents free-ride on active teammates' returns. Simultaneous $\epsilon$-greedy exploration compounds this failure mode through \emph{relative overgeneralisation}~\cite{panait2006biasing,wei2016lenient}: exploratory noise from concurrent teammates obscures the true marginal value of coordinated actions, pushing policies toward low-variance suboptimal equilibria. LAIES~\cite{liu2023laies} tackles agent laziness most closely in target pathology, learning an External States Transition Model to estimate causal influence and injecting intrinsic diligence rewards (IDI and CDI) into the training objective. AnP avoids auxiliary world models and reward shaping entirely, diagnosing laziness online directly from Q-value spreads and empirical action entropy.

\paragraph{Cooperative Multi-Agent Exploration.}
Uniform $\epsilon$-greedy exploration often fails to discover coordinated joint policies in value-factorised MARL. MAVEN~\cite{mahajan2019maven} addresses committed-exploration failure by conditioning joint policies on a hierarchical latent variable trained via mutual-information maximisation. Influence- and curiosity-driven approaches such as EITI/EDTI~\cite{wang2020influence} and EMC~\cite{zheng2021emc} shape rewards using inter-agent mutual information or Q-value prediction errors, while CMAE~\cite{liu2021cmae} restricts exploration to shared bottom-up state subspaces. Unlike methods adding auxiliary networks, intrinsic bonuses, or episodic value estimators, AnP leaves the learner's optimisation objective untouched and resolves exploration interference through structured rank-1 rollout scheduling alone.

\paragraph{Sequential Coordination and Rank-1 Stationarity.}
Decoupling simultaneous agent updates controls multi-agent non-stationarity in well-established ways. On-policy methods such as HAPPO~\cite{kuba2022hatrpo} prove monotonic policy improvement via sequential agent-by-agent updates, while A2PO~\cite{wang2023a2po} demonstrates that update ordering critically impacts sample efficiency. At the rollout level, OVMSE (Hai Zhong et al., 2025)~\cite{zhong2025ovmse} is our closest neighbour, introducing Decentralised Sequential Exploration to perturb one agent at a time ($\epsilon_{\mathrm{dec}} = \epsilon / N$) during offline-to-online fine-tuning. OVMSE relies on a pre-trained offline value anchor (OVM) and selects the exploring agent uniformly at random each step. AnP operates purely online from scratch, targeting the most credit-starved agent diagnostically via $\mathcal{L}_i(t)$, sustaining strictly greedy teammate anchoring ($\epsilon_{-i} = 0$) across a bounded $\phi$-annealed multi-episode pulse, and stabilising post-pulse re-entry with a smooth $\phi$-ramped cooldown. Table~\ref{tab:positioning} summarises these distinctions.

\begin{table}[t]
\centering
\caption{Conceptual comparison of AnP-Auto against representative cooperative MARL, exploration, and sequential-update methods.}
\label{tab:positioning}
\resizebox{\textwidth}{!}{%
\begin{tabular}{lccccc}
\toprule
\textbf{Method} & \textbf{Modifies Loss / Reward} & \textbf{Extra Models} & \textbf{Rollout Stationarity} & \textbf{Diagnostic Targeting} & \textbf{Regime} \\
\midrule
QMIX~\cite{rashid2018qmix} & No & No & No & No & Online \\
MAVEN~\cite{mahajan2019maven} & Yes (MI loss) & Yes (Latent net) & No (Joint latent) & No & Online \\
LAIES~\cite{liu2023laies} & Yes (IDI / CDI) & Yes (ESTM) & No & Implicit (Causal) & Online \\
HAPPO / A2PO~\cite{kuba2022hatrpo,wang2023a2po} & Yes (Seq. PPO) & No & Update-level only & Order-based (A2PO) & On-policy \\
OVMSE~\cite{zhong2025ovmse} & Yes (OVM loss) & Yes (Offline $Q$) & Yes (1-step random) & No (Uniform $1/N$) & Offline-to-Online \\
\midrule
\textbf{AnP-Auto (Ours)} & \textbf{No} & \textbf{No} & \textbf{Yes (Multi-ep $\epsilon_{-i}=0$)} & \textbf{Yes ($\arg\max_i \mathcal{L}_i$)} & \textbf{Pure Online} \\
\bottomrule
\end{tabular}%
}
\end{table}

\section{Methodology}
\label{sec:methodology}

We now formalise \textbf{Anchor and Perturb (AnP)} and its closed-loop automation (\textbf{AnP-Auto}). These are purely online exploration wrappers for value-factorised CTDE learners. Throughout training, QMIX's recurrent agent networks, monotonic mixing hypernetwork, uniform replay sampling, and temporal-difference (TD) loss remain standard; AnP operates strictly by shaping an agent-wise exploration vector at the rollout interface.

\subsection{Preliminaries and Vectorised Rollout Interface}
\label{sec:preliminaries}

We model a fully cooperative multi-agent task as a Decentralised Partially Observable Markov Decision Process (Dec-POMDP). The environment is defined by the tuple $\mathcal{G} = \langle \mathcal{S}, \mathcal{U}, P, r, \mathcal{Z}, O, N, \gamma \rangle$, where $s \in \mathcal{S}$ denotes the true global state. Each agent $i \in \mathcal{N} = \{1, \dots, N\}$ receives a local observation $o_i = O(s, i) \in \mathcal{Z}$ and selects an action $u_i \in \mathcal{A}_i \subseteq \mathcal{U}$ from its available action set. Joint actions $\mathbf{u} = (u_1, \dots, u_N)$ induce state transitions $P(s' \mid s, \mathbf{u})$ and yield a shared team reward $r(s, \mathbf{u})$.

In QMIX~\cite{rashid2018qmix}, each agent conditions a recurrent utility network $Q_i(\tau_i, u_i)$ on its action-observation history $\tau_i$, and a state-conditioned monotonic mixing network $f_{\mathrm{mix}}$ aggregates individual utilities into $Q_{\mathrm{tot}}(\boldsymbol{\tau}, \mathbf{u}; s)$. This satisfies $\partial Q_{\mathrm{tot}} / \partial Q_i \ge 0$. Standard QMIX collects off-policy rollouts using a scalar exploration rate $\epsilon_{\mathrm{nom}}(e)$ shared uniformly across all agents, linearly annealed from $1.0$ to $\epsilon_{\min}$ over $E_{\mathrm{exp}}$ episodes:
\begin{equation}
    \epsilon_{\mathrm{nom}}(e) = \max\!\left(\epsilon_{\min},\; 1.0 - (1.0 - \epsilon_{\min})\frac{e}{E_{\mathrm{exp}}}\right).
    \label{eq:nominal_eps}
\end{equation}

AnP generalises this rollout interface to an agent-wise exploration vector $\boldsymbol{\epsilon}(e) = (\epsilon_1(e), \dots, \epsilon_N(e)) \in [0, 1]^N$. At each timestep, agent $i$ samples a uniformly random valid action from $\mathcal{A}_i$ with probability $\epsilon_i(e)$ and executes $\arg\max_{u \in \mathcal{A}_i} Q_i(\tau_i, u)$ otherwise.

\subsection{Online Model-Free Laziness Diagnosis}
\label{sec:laziness_index}

To detect passive, credit-starved agents online without auxiliary world models, we introduce the \textbf{Laziness Disparity Index} $\mathcal{L}_i(t)$. This is computed over a sliding global-clock window $\mathcal{W}_t$ of the most recent $W$ environment steps collected outside active intervention pulses.

\paragraph{Alive-Masked Action-Value Spread.}
In multi-agent combat benchmarks such as SMAC, eliminated units remain in the environment array with only a placeholder no-op action available. Treating eliminated steps as active decisions would artificially collapse an agent's Q-value variance and action entropy. We therefore maintain a binary alive mask $m_{i,\tau} \in \{0, 1\}$ aligned to a shared global clock across all agents. This sets $m_{i,\tau} = 0$ whenever agent $i$ is inactive (i.e., when only the dead-unit no-op is available) and $m_{i,\tau} = 1$ otherwise. At every alive step $\tau \in \mathcal{W}_t$, we compute the instantaneous Q-value spread over the valid action set $\mathcal{A}_{i,\tau}$:
\begin{equation}
    s_{i,\tau} =
    \begin{cases}
        \max\limits_{u \in \mathcal{A}_{i,\tau}} Q_i(\tau_i, u) - \dfrac{1}{|\mathcal{A}_{i,\tau}|} \sum\limits_{u \in \mathcal{A}_{i,\tau}} Q_i(\tau_i, u), & |\mathcal{A}_{i,\tau}| > 1, \\[10pt]
        0, & \text{otherwise.}
    \end{cases}
    \label{eq:q_spread}
\end{equation}
We aggregate each agent's sustained decision sensitivity across the full global window $W$:
\begin{equation}
    c_i(t) = \frac{1}{W} \sum_{\tau \in \mathcal{W}_t} m_{i,\tau}\, s_{i,\tau}.
    \label{eq:window_spread}
\end{equation}

Normalising by $W$ rather than each agent's individual alive count ensures that units which either contribute flat utility spreads while alive or disengage early register low $c_i(t)$ relative to the team's most decisive contributor $\max_j c_j(t)$.

\paragraph{Empirical Alive Action Entropy.}
Low Q-spread alone can arise transiently when an agent faces genuinely symmetric choices. To isolate collapsed, repetitive behaviour, we compute each agent's empirical action entropy strictly over its $n_i^{\mathrm{alive}} = \sum_{\tau \in \mathcal{W}_t} m_{i,\tau}$ active steps in $\mathcal{W}_t$:
\begin{equation}
    p_i(u) = \frac{1}{n_i^{\mathrm{alive}}} \sum_{\tau \in \mathcal{W}_t} m_{i,\tau}\,\mathbb{I}[u_{i,\tau} = u],
    \qquad
    H_i(t) = -\sum_{u : p_i(u) > 0} p_i(u)\ln p_i(u),
    \label{eq:action_entropy}
\end{equation}
with $c_i(t) = 0$ and $H_i(t) = 0$ whenever $n_i^{\mathrm{alive}} = 0$.

\paragraph{Laziness Disparity Index.}
Combining relative Q-spread suppression with exponential entropy damping yields:
\begin{equation}
    \mathcal{L}_i(t) = \left(1 - \frac{c_i(t)}{\max_{j \in \mathcal{N}} c_j(t) + \delta}\right) \exp\!\big(-\beta H_i(t)\big),
    \label{eq:laziness_index}
\end{equation}
where $\delta = 10^{-5}$ prevents division by zero and $\beta > 0$ controls entropy sensitivity (we fix $\beta = 1.0$). An agent receives a high $\mathcal{L}_i(t)$ only when its Q-value sensitivity is sharply suppressed relative to the team leader and its executed policy has collapsed into low-entropy repetition. In rich action spaces such as 10-agent SMAC \texttt{MMM}, active units naturally retain moderate tactical entropy ($H_i \approx 1.5\text{--}2.2$ nats across movement and attack targets), compressing $\mathcal{L}_i(t)$ into a compact band ($\approx 0.05\text{--}0.15$) in which persistently lagging agents cleanly separate above a calibrated threshold $\tau_L$.

\subsection{Closed-Loop Rank-1 Perturbation and Smooth Un-Anchoring}
\label{sec:anp_schedule}

AnP-Auto governs the rollout vector $\boldsymbol{\epsilon}(e)$ through a three-phase state machine (Nominal, Rank-1 Pulse, and Smooth Cooldown), controlled by a performance-gated trigger.

\paragraph{Gated Automated Triggering.}
At the end of episode $e$, once the sliding probe buffer is full ($|\mathcal{W}_t| = W$), AnP-Auto checks five conditions before initiating an intervention: (i) training has passed an early-exploration borderline ($e \ge E_{\mathrm{border}}$); this prevents premature interference while joint policies are still forming; (ii) neither a pulse nor a post-pulse cooldown is active; (iii) the cumulative pulse count remains below a budget ($k < K$); (iv) the latest greedy evaluation win rate $\bar{w}(e)$ lies below a mastery ceiling ($w_{\mathrm{ceil}} = 0.80$), avoiding disruption to already solved policies; and (v) the worst-off agent exceeds the laziness threshold:
\begin{equation}
    i^* = \arg\max_{i \in \mathcal{N}} \mathcal{L}_i(t),
    \qquad
    \mathcal{L}_{i^*}(t) > \tau_L.
    \label{eq:trigger}
\end{equation}

When all five conditions hold, AnP-Auto schedules a bounded intervention window $[e_{\mathrm{start}}, e_{\mathrm{end}})$ with $e_{\mathrm{start}} = e + 1$ and $e_{\mathrm{end}} = \min(e_{\mathrm{start}} + D_p, E_{\mathrm{tot}})$. We also increment $k \leftarrow k + 1$.

\paragraph{Phase 1: Rank-1 $\phi$-Annealed Perturbation ($e \in [e_{\mathrm{start}}, e_{\mathrm{end}})$).}
During an active pulse, AnP isolates agent $i^*$ ($M = 1$) while strictly anchoring all $N - 1$ teammates to deterministic greedy execution:
\begin{equation}
    \epsilon_i(e) =
    \begin{cases}
        \epsilon_{\min} + \Delta\epsilon \cdot (1 - u)^{\phi}, & i = i^*, \\
        0, & i \neq i^*,
    \end{cases}
    \qquad
    u = \frac{e - e_{\mathrm{start}}}{\max(1,\, e_{\mathrm{end}} - e_{\mathrm{start}})} \in [0, 1),
    \label{eq:pulse_schedule}
\end{equation}
where $\Delta\epsilon = 0.25$ (yielding an initial probe peak of $\epsilon_{\min} + \Delta\epsilon = 0.30$ for $\epsilon_{\min} = 0.05$) and $\phi > 1$ (fixed at $\phi = 1.5$) is a non-linear shaping exponent. Choosing $\phi > 1$ front-loads exploratory variance at the start of the pulse to escape the lazy local optimum, then tapers smoothly toward $\epsilon_{\min}$ so QMIX can consolidate improved transitions before the pulse ends. Probe updates to $\mathcal{W}_t$ pause throughout the pulse so injected exploration noise does not contaminate laziness estimates.

\paragraph{Phase 2: Smooth $\phi$-Ramped Un-Anchoring ($e \in [e_{\mathrm{end}}, e_{\mathrm{end}} + D_c)$).}
Abruptly restoring nominal exploration across all $N - 1$ anchored teammates at $e_{\mathrm{end}}$ can introduce sudden distributional shock that overwrites newly discovered coordination patterns in the replay buffer. Upon pulse completion, AnP flushes $\mathcal{W}_t$ to clear pre-pulse statistics, resumes probe logging, and smoothly un-anchors teammates over $D_c$ cooldown episodes using the dual $\phi$-shaped ramp:
\begin{equation}
    \epsilon_i(e) =
    \begin{cases}
        \epsilon_{\mathrm{nom}}(e), & i = i^*, \\
        \epsilon_{\mathrm{nom}}(e) \cdot v^{\phi}, & i \neq i^*,
    \end{cases}
    \qquad
    v = \frac{e - e_{\mathrm{end}}}{\max(1,\, D_c)} \in [0, 1).
    \label{eq:cooldown_schedule}
\end{equation}

Because $\phi > 1$ gives $\frac{d}{dv}(v^{\phi})\big|_{v=0} = 0$, teammates remain near-deterministic immediately after release while QMIX propagates updated Bellman targets through the replay buffer. Automated re-triggering remains blocked during cooldown. Once $e \ge e_{\mathrm{end}} + D_c$, all agents return to the nominal schedule $\epsilon_i(e) = \epsilon_{\mathrm{nom}}(e)$.

\subsection{Stationary Projection Intuition}
\label{sec:intuition}

From the perspective of the targeted agent $i^*$, uniform $\epsilon$-greedy exploration embeds its learning problem in a non-stationary environment whose effective transition kernel shifts whenever any teammate $j \neq i^*$ takes a random action. Because the probability that all $N - 1$ teammates act greedily on a single step is $(1 - \epsilon_{\mathrm{nom}})^{N-1}$, even modest per-agent noise compounds over multi-step episodes in large teams (e.g., $N = 10$ on \texttt{MMM}).

Enforcing $\epsilon_{-i^*} = 0$ throughout the pulse projects the joint Dec-POMDP onto an approximately stationary single-agent MDP conditioned on the fixed greedy background policy $\boldsymbol{\pi}_{-i^*}^{\mathrm{greedy}}$. Under QMIX's monotonic mixing constraint ($\partial Q_{\mathrm{tot}} / \partial Q_{i^*} \ge 0$), any consistent team-return gain discovered by $i^*$ against this stationary background aligns directly with positive TD advantage updates to $Q_{i^*}$, restoring credit assignment without modifying the optimisation objective.

\begin{algorithm}[t]
\caption{Closed-Loop Anchor and Perturb (AnP-Auto) for QMIX}
\label{alg:anp_auto}
\begin{algorithmic}[1]
\Require QMIX learner, replay buffer $\mathcal{B}$, probe window $W$, threshold $\tau_L$, borderline $E_{\mathrm{border}}$, ceiling $w_{\mathrm{ceil}}$, pulse duration $D_p$, cooldown $D_c$, max pulses $K$, shaping exponent $\phi > 1$.
\State Initialise $k \leftarrow 0$, $i^* \leftarrow \mathrm{None}$, $\mathcal{W} \leftarrow \emptyset$, $\bar{w} \leftarrow 0$.
\For{episode $e = 0, \dots, E_{\mathrm{tot}} - 1$}
    \State Compute nominal rate $\epsilon_{\mathrm{nom}}(e)$ via Eq.~\eqref{eq:nominal_eps}.
    \If{$k > 0$ \textbf{and} $e \in [e_{\mathrm{start}}, e_{\mathrm{end}})$} \Comment{Rank-1 Perturbation Pulse}
        \State Set $\boldsymbol{\epsilon}(e)$ via Eq.~\eqref{eq:pulse_schedule} with $\epsilon_{i^*}(e) > \epsilon_{\min}$ and $\epsilon_{-i^*}(e) = 0$; pause probe logging.
    \ElsIf{$k > 0$ \textbf{and} $e \in [e_{\mathrm{end}}, e_{\mathrm{end}} + D_c)$} \Comment{Smooth Un-Anchoring Cooldown}
        \State On entry, flush $\mathcal{W}$, resume probe logging, and store $i_{\mathrm{last}} \leftarrow i^*$.
        \State Set $\boldsymbol{\epsilon}(e)$ via Eq.~\eqref{eq:cooldown_schedule} (ramping $j \neq i_{\mathrm{last}}$ via $v^\phi$).
    \Else \Comment{Nominal Phase}
        \State Set $\epsilon_i(e) \leftarrow \epsilon_{\mathrm{nom}}(e)$ for all $i \in \mathcal{N}$.
    \EndIf
    \State Roll out episode under $\boldsymbol{\epsilon}(e)$, appending alive-masked $(u_{i,\tau}, s_{i,\tau})$ to $\mathcal{W}$ outside pulses.
    \State Insert episode into $\mathcal{B}$ and perform one standard QMIX TD update on a uniform minibatch.
    \If{$e \ge E_{\mathrm{border}}$ \textbf{and} not in pulse/cooldown \textbf{and} $k < K$ \textbf{and} $|\mathcal{W}| = W$}
        \State Compute $\mathcal{L}_i(t)$ for all $i \in \mathcal{N}$ via Eqs.~\eqref{eq:q_spread}--\eqref{eq:laziness_index}; set $j^* \leftarrow \arg\max_i \mathcal{L}_i(t)$.
        \If{$\mathcal{L}_{j^*}(t) > \tau_L$ \textbf{and} $\bar{w} < w_{\mathrm{ceil}}$}
            \State Set $i^* \leftarrow j^*$, $e_{\mathrm{start}} \leftarrow e + 1$, $e_{\mathrm{end}} \leftarrow \min(e_{\mathrm{start}} + D_p, E_{\mathrm{tot}})$, $k \leftarrow k + 1$.
        \EndIf
    \EndIf
    \State Periodically update $\bar{w}$ across greedy evaluation episodes ($\boldsymbol{\epsilon} = \mathbf{0}$).
\EndFor
\end{algorithmic}
\end{algorithm}

\section{Experiments}
\label{sec:experiments}

We evaluate AnP-Auto on the heterogeneous 10-agent StarCraft Multi-Agent Challenge (SMAC) \texttt{MMM} scenario against vanilla QMIX across five paired random seeds. Our experiments address three questions: whether purely online rank-1 rollout scheduling improves terminal win rates and returns without modifying QMIX's loss function or architecture, how policies evolve through perturbation, cooldown, and consolidation phases, and whether the alive-masked Laziness Disparity Index $\mathcal{L}_i(t)$ reliably identifies credit-starved agents in practice.

\subsection{Experimental Setup}
\label{sec:exp_setup}

SMAC \texttt{MMM} combines Medivac support with Marauder and Marine combat units against an identical 10-unit enemy force. Both methods employ the same QMIX backbone, trained with learning rate $5 \times 10^{-4}$, minibatch size $32$, uniform episode replay buffer capacity of $2{,}000$ episodes, one gradient update per episode, and target network synchronisation every $200$ updates. Nominal $\epsilon$-greedy exploration anneals linearly from $1.0$ to $\epsilon_{\min} = 0.05$ over $E_{\mathrm{exp}} = 1{,}000$ episodes.

To isolate rollout scheduling effects cleanly, we disable all auxiliary replay modifications (no prioritised or elite reservoir sampling) and maintain identical TD loss, mixing network structure, and environment rewards to vanilla QMIX. Across AnP-Auto runs, configuration parameters are: probe window $W = 1{,}000$ steps, entropy damping $\beta = 1.0$, mastery ceiling $w_{\mathrm{ceil}} = 0.80$, pulse duration $D_p = 200$ episodes, smooth un-anchoring cooldown $D_c = 200$ episodes, pulse budget $K = 1$, perturbation amplitude $\Delta\epsilon = 0.25$ (peak $\epsilon = 0.30$), and shaping exponent $\phi = 1.5$.

Every $25$ training episodes, we evaluate the current greedy policy ($\boldsymbol{\epsilon} = \mathbf{0}$) across $20$ test episodes, recording mean win rate and mean return. We compare five paired seeds ($\{7, 15, 25, 36, 42\}$). Four seeds ($\{7, 25, 36, 42\}$) run for a uniform $4{,}500$-episode horizon; Seed~15 originates from an earlier $4{,}000$-episode protocol with an earlier trigger borderline. On four of the five seeds ($7, 15, 25, 36$), vanilla QMIX and AnP-Auto follow identical evaluation trajectories prior to pulse onset ($\max|\Delta w_{\mathrm{pre}}| = 0.00$), providing exact counterfactual baselines. On Seed~42, minor simulator non-determinism caused pre-pulse trajectory divergence starting at Episode~1750 ($1{,}300$ episodes before its pulse). We retain Seed~42 for completeness and report aggregate statistics over both the full 5-seed cohort and the uniform $4{,}500$-episode 4-seed cohort.

\subsection{Main Results on SMAC \textnormal{\texttt{MMM}}}
\label{sec:main_results}

Table~\ref{tab:mmm_main} summarises terminal performance averaged over the final $500$ training episodes ($20$ evaluation checkpoints, or $400$ greedy test episodes per seed per method). AnP-Auto outperforms vanilla QMIX on all five seeds across both greedy evaluation win rate and evaluation return, as well as terminal training return.

Across all five seeds, AnP-Auto raises terminal mean win rate from $30.20\% \pm 4.28\%$ to $37.20\% \pm 4.55\%$, an absolute gain of $+7.00 \pm 4.29$ percentage points ($+23.2\%$ relative improvement; two-sided paired $t(4) = 3.65, p = 0.0217$). Terminal mean evaluation return increases from $12.56 \pm 0.45$ to $13.27 \pm 0.29$ ($+0.71 \pm 0.26$; paired $t(4) = 6.08, p = 0.0037$). Restricting to the uniform $4{,}500$-episode 4-seed cohort widens the margin to $+7.69$ percentage points in win rate ($30.94\% \to 38.62\%$; paired $t(3) = 3.33, p = 0.0448$) and $+0.73$ in evaluation return ($12.59 \to 13.32$; paired $t(3) = 4.99, p = 0.0155$). Gains are largest on the weakest vanilla baseline run (Seed~25: $+13.75$ percentage points), confirming that rank-1 perturbations specifically rescue runs trapped in suboptimal equilibria.

\begin{table}[t]
\centering
\caption{Paired counterfactual evaluation on SMAC \texttt{MMM} over the terminal $500$ episodes ($20$ checkpoints, $400$ greedy test episodes per seed). Logged pulse windows reflect $25$-episode checkpoint intervals spanning each $200$-episode rank-1 pulse. Statistical significance is evaluated via two-sided paired $t$-tests.}
\label{tab:mmm_main}
\resizebox{\textwidth}{!}{%
\begin{tabular}{ccccccccc}
\toprule
& & & \multicolumn{3}{c}{\textbf{Terminal Eval Win Rate (\%)}} & \multicolumn{3}{c}{\textbf{Terminal Eval Return}} \\
\cmidrule(lr){4-6} \cmidrule(lr){7-9}
\textbf{Seed} & \textbf{Horizon} & \textbf{Pulse Window} & \textbf{Vanilla} & \textbf{AnP-Auto} & \textbf{$\Delta$ (pp)} & \textbf{Vanilla} & \textbf{AnP-Auto} & \textbf{$\Delta$} \\
\midrule
7  & 4{,}500 & 3700--3875 & 34.75 & \textbf{42.00} & +7.25  & 13.18 & \textbf{13.75} & +0.57 \\
15 & 4{,}000 & 2025--2200 & 27.25 & \textbf{31.50} & +4.25  & 12.47 & \textbf{13.08} & +0.61 \\
25 & 4{,}500 & 2825--3000 & 24.50 & \textbf{38.25} & +13.75 & 11.92 & \textbf{13.07} & +1.16 \\
36 & 4{,}500 & 3025--3200 & 33.50 & \textbf{40.75} & +7.25  & 12.64 & \textbf{13.32} & +0.68 \\
42 & 4{,}500 & 3050--3225 & 31.00 & \textbf{33.50} & +2.50  & 12.61 & \textbf{13.12} & +0.51 \\
\midrule
\multicolumn{3}{l}{\textbf{All 5 Seeds (Mean $\pm$ SD)}} & 30.20 $\pm$ 4.28 & \textbf{37.20 $\pm$ 4.55} & \textbf{+7.00} & 12.56 $\pm$ 0.45 & \textbf{13.27 $\pm$ 0.29} & \textbf{+0.71} \\
\multicolumn{3}{l}{\emph{Paired $t$-test ($df=4$)}} & \multicolumn{3}{c}{$t = 3.65,\; p = 0.0217$} & \multicolumn{3}{c}{$t = 6.08,\; p = 0.0037$} \\
\midrule
\multicolumn{3}{l}{\textbf{4{,}500-Ep Cohort ($N=4$)}} & 30.94 $\pm$ 4.57 & \textbf{38.62 $\pm$ 3.76} & \textbf{+7.69} & 12.59 $\pm$ 0.52 & \textbf{13.32 $\pm$ 0.31} & \textbf{+0.73} \\
\multicolumn{3}{l}{\emph{Paired $t$-test ($df=3$)}} & \multicolumn{3}{c}{$t = 3.33,\; p = 0.0448$} & \multicolumn{3}{c}{$t = 4.99,\; p = 0.0155$} \\
\bottomrule
\end{tabular}%
}
\end{table}

\subsection{Training Dynamics: Transient Dip and Late-Stage Escape}
\label{sec:dynamics}

Table~\ref{tab:phase_bins} partitions greedy evaluation win rate across six chronological intervals for the uniform $4{,}500$-episode cohort ($\{7, 25, 36, 42\}$). AnP-Auto follows a distinct two-stage trajectory rather than producing an immediate mid-training improvement. During Episodes~1--2000, both methods track almost identically ($3.12\%$ and $\approx 12.4\%$). Across Episodes~2001--4000, as automated pulses fire between Episodes~2801 and~3680 and perturbed rollouts flush into the $2{,}000$-episode replay buffer, AnP-Auto temporarily trails vanilla QMIX, reaching a $-4.88$ percentage-point trough during Episodes~3501--4000 ($31.31\%$ vs.\ $36.19\%$). This temporary lag reflects the expected consolidation cost of replacing stale nominal transitions with perturbed single-agent trajectories.

In the terminal window (Episodes~4001--4500), however, the two trajectories move in opposite directions. Vanilla QMIX regresses by $-5.25$ percentage points ($36.19\% \to 30.94\%$), showing late-stage value drift under uniform $\epsilon_{\min} = 0.05$ noise, whereas AnP-Auto climbs by $+7.31$ percentage points ($31.31\% \to 38.62\%$) while reducing cross-seed variance (SD $3.76\%$ vs.\ $4.57\%$). AnP-Auto thus trades a brief, bounded exploration dip for escaping late-stage QMIX stagnation and regression.

\begin{table}[t]
\centering
\caption{Chronological greedy evaluation win rate (Mean $\pm$ SD, \%) across the uniform $4{,}500$-episode cohort (Seeds~7, 25, 36, 42).}
\label{tab:phase_bins}
\begin{tabular}{lccc}
\toprule
\textbf{Episode Range} & \textbf{Vanilla QMIX (\%)} & \textbf{AnP-Auto (\%)} & \textbf{$\Delta$ (pp)} \\
\midrule
1--1000    &  3.12 $\pm$  4.52 &  3.12 $\pm$  4.52 &  +0.00 \\
1001--2000 & 12.34 $\pm$ 10.87 & 12.44 $\pm$ 11.05 &  +0.09 \\
2001--3000 & 24.03 $\pm$  8.61 & 23.09 $\pm$  7.45 &  $-0.94$ \\
3001--3500 & 26.12 $\pm$  8.45 & 24.50 $\pm$  6.55 &  $-1.62$ \\
3501--4000 & 36.19 $\pm$  7.70 & 31.31 $\pm$  5.75 &  $-4.88$ \\
4001--4500 & 30.94 $\pm$  4.57 & \textbf{38.62 $\pm$ 3.76} & \textbf{+7.69} \\
\bottomrule
\end{tabular}
\end{table}

\subsection{Diagnostic Case Study: Tracking \texorpdfstring{$\mathcal{L}_i(t)$}{Li(t)} on Seed~25}
\label{sec:diagnostic_case}

Full per-checkpoint probe telemetry ($\max_i \mathcal{L}_i(t)$ and $\arg\max_i \mathcal{L}_i(t)$) was recorded on Seed~25 to examine closed-loop diagnostic behaviour during training. Across all $33$ consecutive evaluation checkpoints from Episode~2000 through Episode~2800 prior to intervention, the probe flags \textbf{Agent~7} every time ($33/33$ checkpoints), demonstrating that $\mathcal{L}_i(t)$ locks onto a stable, role-level credit-assignment bottleneck rather than transient stochastic noise. At Episode~2800, $\mathcal{L}_7(t)$ reaches $0.1035$, triggering the $200$-episode rank-1 pulse on Agent~7 (Episodes~2801--3000).

At the first cooldown checkpoint (Episode~3025), the probe resets ($\mathcal{L} = 0.0$ while refilling); once the $1{,}000$-step window refills over Episodes~3050--3100, $\max_i \mathcal{L}_i(t)$ drops to a mean of $0.0703$, accompanied by an immediate cooldown win-rate surge to $34.38\%$ compared with $20.63\%$ for vanilla QMIX ($+13.75$ percentage points). Late in training (Episodes~4025--4500), Agent~7 again registers the highest $\mathcal{L}_i(t)$ (mean $0.0984$, peaking at $0.1526$), yet AnP-Auto sustains a $+13.75$ percentage-point win-rate lead over vanilla QMIX ($38.25\%$ vs.\ $24.50\%$). Because structural role asymmetry in heterogeneous teams like \texttt{MMM} naturally produces persistent Q-spread differences relative to the team's most decisive unit ($\max_j c_j$), AnP-Auto does not need to drive $\mathcal{L}_i(t)$ permanently to zero; a single bounded stationary probe ($K = 1$) suffices to move the stalled agent into a superior joint policy basin without repeatedly destabilising late-stage execution.

\section{Discussion, Limitations, and Future Work}
\label{sec:discussion}

Our \texttt{MMM} results demonstrate that late-stage QMIX stagnation does not stem solely from monotonic mixing expressivity; a substantial portion arises from exploration-induced non-stationarity during rollout collection. Holding $N - 1$ teammates strictly greedy ($\epsilon_{-i} = 0$) over a bounded 200-episode window temporarily restores single-agent stationarity, allowing QMIX's unmodified TD backups to re-estimate the marginal contribution of a stalled unit. Evaluating AnP prematurely during its mid-training dip (Episodes~3501--4000) would mask gains that emerge only as vanilla QMIX begins drifting downward; perturbed rollouts must displace stale nominal transitions in the replay buffer before consolidation can occur.

Telemetry from Seed~25 clarifies an important property of model-free laziness heuristics on heterogeneous teams. Support and frontline units naturally exhibit different baseline Q-value spreads relative to $\max_j c_j$, so $\mathcal{L}_i(t)$ remains nonzero even after a unit recovers. Bounding the pulse budget ($K$) and gating interventions on a mastery ceiling ($w_{\mathrm{ceil}}$) are therefore essential to prevent repeated disruption of specialised roles. The entropy damping term $\exp(-\beta H_i)$ scales with the cardinality of each agent's valid action space, meaning the trigger threshold $\tau_L$ requires domain-specific calibration ($\tau_L \approx 0.10\text{--}0.11$ on 10-agent \texttt{MMM}). Normalising empirical entropy by $\ln |\mathcal{A}_i|$ or adopting a running $z$-score threshold would render $\tau_L$ invariant across map sizes—a straightforward refinement for broader applicability.

Several experimental extensions remain open. The present validation focuses on paired counterfactual runs of QMIX on SMAC \texttt{MMM} with a single pulse budget ($K = 1$) and uniform replay sampling, while noting minor protocol heterogeneity across seeds: Seed~15 terminates at $4{,}000$ episodes, and Seed~42 exhibits pre-pulse simulator divergence starting at Episode~1750. Immediate next steps include expanding evaluation to additional SMAC maps (such as \texttt{3s5z} and \texttt{corridor}), isolating each component through ablations over $\phi$, cooldown ramping, multi-pulse budgets ($K \ge 2$), and selective elite replay. Verifying orthogonality on expressive mixers such as Weighted QMIX~\cite{rashid2020wqmix} and QPLEX~\cite{wang2021qplex} is also warranted. Finally, in strictly sparse-reward domains where early Q-value spreads remain flat across all agents, passive heuristics lose signal; coupling causal diligence estimation inspired by LAIES~\cite{liu2023laies} with AnP's rank-1 anchored rollouts forms the focus of our follow-up framework, \textbf{IExPert}.

\section{Conclusion}
\label{sec:conclusion}

We introduced \textbf{Anchor and Perturb (AnP)} and its closed-loop automation (\textbf{AnP-Auto}), a lightweight, purely online rollout-scheduling wrapper that mitigates lazy-agent stagnation in value-factorised MARL without modifying the underlying learner's architecture, loss function, or reward signal. By pairing an alive-masked Laziness Disparity Index $\mathcal{L}_i(t)$ with performance-gated rank-1 $\phi$-annealed exploration pulses ($\epsilon_{-i} = 0$) and smooth $\phi$-ramped un-anchoring, AnP-Auto converts noisy joint exploration into targeted stationary single-agent probes. Across five paired random seeds on SMAC \texttt{MMM}, AnP-Auto overcomes late-stage vanilla QMIX regression and improves terminal 500-episode greedy win rate on every seed by an average of $+7.00$ percentage points ($p = 0.0217$), alongside a $+0.71$ gain in evaluation return ($p = 0.0037$). These results demonstrate that structured rollout-level stationarity constitutes a simple, orthogonal lever for unlocking latent coordination in cooperative MARL.

\bibliographystyle{plainnat}
\bibliography{references}

\end{document}